\documentclass[showpacs
,showkeys
,nofootinbib
,amsfonts,amssymb
,floatfix
,twocolumn,tightenlines
,groupaddresses
,prc
,longbibliography
]{revtex4-2}
\pdfoutput=1

\usepackage{graphicx}  
\usepackage{mathtools}
\usepackage{amsmath, amssymb, bm,bbm}   
\usepackage{xcolor}
\usepackage{natbib}
\usepackage{hyperref}

\usepackage{longtable}
\usepackage{csvsimple} 
\usepackage{adjustbox} 

\usepackage[normalem]{ulem} 

\usepackage{ifthen}

\usepackage{siunitx} 
\usepackage{balance}

\usepackage{orcidlink}

\newcommand*\xbar[1]{%
   \hbox{%
     \vbox{%
       \hrule height 0.5pt 
       \kern0.3ex
       \hbox{%
         \kern-0.1em
         \ensuremath{#1}%
         \kern-0.1em
       }%
     }%
   }%
}

\begin{document}

\title{Proton-proton scattering on a quantum computer }

\author{%
 Ratna Khadka\,\orcidlink{0009-0009-6628-7028}}
\email{rk973@.msstate.edu}
\author{%
Gautam Rupak\,\orcidlink{0000-0001-6683-177X}}
\email{grupak@ccs.msstate.edu}

\affiliation{Department of Physics \& Astronomy and HPC$^2$ Center for 
Computational Sciences, Mississippi State
University, Mississippi State, MS 39762, USA}

\begin{abstract}
Scattering of charged particles is ubiquitous in nuclear physics. We calculate the proton-proton $s$-wave phase shift at low energy relevant to solar physics. The phase shift is calculated from the ratio of the regular and irregular solutions to the radial Schr\"odinger equation on a hard spherical wall boundary for the ground state. The ground state energy is calculated using a hybrid quantum-classical variational algorithm. A theory with short-ranged nuclear interaction in the presence of the long-ranged Coulomb force is used to describe the scattering. The theory is discretized on a spatial lattice for adaptation to the quantum computer in the second quantized language. The phase shifts at low momenta are accurately reproduced.   
\end{abstract}

\keywords{proton-proton scattering, Coulomb interaction, Quantum Computer}
\maketitle


Proton-proton fusion initiates the chain of reactions, converting four protons into a helium-4 nucleus, that powers the Sun. Following the seminal work by Bethe and Critchfield on proton-proton fusion~\cite{BetheCritchfield:1938}, 
Bahcall and May~\cite{BahcallMay:1969} set the benchmark for future evaluations of this important reaction, see Ref.~\cite{Acharya:2024} and references therein.  Solar models are constrained by this reaction which is sensitive to solar properties such as the central density and temperature. Measurements of neutrinos from solar reactions are crucial for the current understanding of physics beyond the solar model~\cite{Ahmad:2001an,Ahmad:2002jz,Bahcall:1989}. Among the important solar reactions, proton fusion is one that cannot be directly measured in laboratories. Theoretical calculations are the only means available for this reaction rate at solar energies~\cite{Acharya:2024}.  

Effective field theory (EFT) methods, where the short-ranged nuclear interactions are approximated from low-energy symmetries, provide the most accurate values for the proton-fusion reaction rate~\cite{Acharya:2024}. At the solar energies, an EFT where the nuclear interactions are approximated by zero-ranged forces~\cite{Bedaque:1997qi, Chen:1999tn} provides a convenient and accurate framework for the calculations~\cite{Acharya:2024,Kong:2021, Butler:2001}.  An important ingredient in the fusion rate calculation is the initial state proton-proton scattering in the presence of both the short-ranged nuclear and the long-ranged Coulomb force. The EFT for this was presented in Refs.~\cite{Kong-Randal:1999,Kong-Randal:1999B, Kong:2021} that we adopt for our calculation of the proton-proton elastic scattering on a quantum computer.    

A typical many-body calculation on a classical computer requires Monte Carlo simulations in Euclidean times. These suffer from the so-called fermionic sign problem~\cite{Loh:1990} that results in a diminished signal-to-noise ratio as the system size grows. Further, real-time correlation function calculations from 
Euclidean time simulations are complicated due to the Maiani-Testa~\cite{Maiani:1990ca} no-go theorem. Quantum computations directly perform unitary evolution in real time, and avoids the fermionic sign problem. There is an expectation that quantum computation would allow calculations of dynamical properties in many-body systems in a way that is not feasible in classical computation. However, quantum computation of physical observables would require a new paradigm in computations. Many of the classical methods of computations are not feasible on a quantum computer. For example, attempts at proton fusion using unitary evolution would result in a Rabi oscillation between the forward and backward reactions on a quantum computer. An algorithm for calculating the transition rates of inelastic reactions using the Rabi oscillation frequency has been proposed~\cite{Bedaque:2022ftd}. Recently, there has been several comparable papers on calculating elastic scattering of neutral particles on quantum computers~\cite{Sharma:2023,Wang:2024,Turro:2024ksf}. In this work, we consider elastic scattering of charged particles---two protons.

The $s$-wave scattering phase shift is calculated using a spherical wall boundary condition on the relative separation of the protons~\cite{Carlson:1984,Borasoy:2007} that relates the phase shift as a function of momentum to the ratio of the regular and irregular Coulomb wave functions. A similar approach was taken in Ref.~\cite{Sharma:2023} except we differ in the method used in the calculation of the momentum, which is similar to the work in Ref.~\cite{Yusf:2024igb}. Coincidentally, Refs.~\cite{Yusf:2024igb} and ~\cite{Wang:2024} both use 
the Busch-Englert-Rza\.zewski-Wilkens (BERW) formula~\cite{Busch:1998cey} for the phase shift though they again differ in the calculation of the momentum. The BERW formula has been extended to include scattering of charged particles~\cite{Guo:2022trap,Zhang:2024cc,Zhang:2024scatter}, though we apply the spherical wall method that has been successful in classical lattice EFT calculations for proton fusion~\cite{Rupak:2014xza}, proton-deuteron scattering~\cite{Elhatisari:2016hby} and alpha-alpha scattering~\cite{Elhatisari:2015iga}. 

In section~\ref{sec:scattering}, we define  proton-proton scattering at low momentum with a non-relativistic Schr\"odinger equation in relative coordinates, which can be formulated on a 3-dimensional spatial lattice for numerical computation as shown  in Refs.~\cite{RaviThesis,Rupak:2014xza}. Here we take a different approach and perform a partial-wave expansion. The formal expressions for $s$-wave Coulomb scattering phase shift is presented in this section.   The partial-wave expansion leads to a 1-dimensional problem in radial relative coordinate, which makes the quantum computation simpler. The formulation of the scattering problem for quantum computation is done in section~\ref{sec:qc}. The phase shift calculation requires calculation of the low energy spectrum which is done with a hybrid quantum-classical variational algorithm that we present in subsection~\ref{subsec:relaxation}. The results of the calculations are presented in section~\ref{sec:results}. We present results from both noisy simulations and a physical quantum processor unit (QPU). We conclude with some final thoughts in section~\ref{sec:conclusions}.

\section{2-particle scattering}
\label{sec:scattering}
In the center-of-mass (cm) coordinates, the 2-particle Schr{\"o}dinger equation reads:
\begin{align}\label{eq:3D_hamiltonian}
    \left[-\frac{1}{2\mu}\bm\nabla^2+V(r)\right]\psi(\bm{r})=E\psi(\bm{r})\,,
\end{align}
with reduced mass $\mu= \SI{469.15}{\mega\eV}$ for protons in natural units $\hbar=1=c$. The partial-wave decomposition in spherical coordinates $\psi(\bm{r})=\psi(r,\theta,\phi)= Y_l^m(\theta,\phi) u(r)/r$, gives 
\begin{align}\label{eq:radial_hamiltonian}
   \left[ -\frac{1}{2\mu}\frac{d^2}{dr^2}+ V(r)+\frac{1}{2\mu}\frac{l(l+1)}{r^2}\right]u(r) = E u(r)\,,
\end{align}
where the angular momentum eigenvalues $l=0, 1,\cdots$,  correspond to $s$-, $p$-, and higher partial waves, respectively. This form of the 1-dimensional Schr\"odinger equation is convenient for quantum computations.
The interaction potential between the protons $V(r)$ is comprised of a short-ranged nuclear $V_s(r)$ and a long-ranged Coulomb $V_c(r)=\alpha_{EM}/r$ part with the fine-structure constant $\alpha_{EM}=1/137$. There is a wall at $r=0$ to impose the boundary condition  $u(0)=0$. 

The elastic scattering amplitude for incoming-outgoing cm momenta $\bm{p}$-$\bm{p}'$ is conventionally written as 
\begin{align}
    T(E;\bm{p}',\bm{p})=T_c(E;\bm{p}',\bm{p})+T_{sc}(E;\bm{p}',\bm{p})\,,
\end{align}
with $|\bm{p}|=p =|\bm{p}'|$ and cm energy $E=p^2/(2\mu)$. The purely Coulomb contribution is expanded in partial waves with Legendre polynomials $P_l(x)$ such that  
\begin{multline}
T_c(E;\bm{p}',\bm{p})=\sum_{l=0}^\infty(2l+1)T^{(l)}_c(E;p)P_l(\hat{\bm p}'\cdot\bm{p})\\
\equiv-\frac{2\pi}{\mu}\sum_{l=0}^\infty(2l+1) \frac{e^{i2\sigma_l}-1}{2ip}P_l(\hat{\bm p}'\cdot\bm{p})\,,
\end{multline}
where the Coulomb phase shifts are known analytically as $\sigma_l(p)= \operatorname{arg}\Gamma(l+1+i\eta_p)$ with the Sommerfeld parameter $\eta_p=\alpha_{EM}\mu/p$. 

The so called Coulomb-subtracted elastic amplitude $T_{sc}(E;\bm{p}',\bm{p})$ can also be expanded in partial waves with
\begin{align}
   T_{sc}^{(l)}(E;p) =-\frac{2\pi}{\mu}\frac{e^{i2\sigma_l}}{p\cot\delta_l-ip}\,,
\end{align}
where $\delta_l(p)$ is referred to as the Coulomb-subtracted phase shift with the full phase shift given simply by $\delta_l(p)+\sigma_l(p)$ for each partial wave $l$. 

At low momentum, the phase shift $\delta_l(p)$ is known to be an analytic function of the energy $E$ for short-ranged interactions. In the presence of the long-ranged Coulomb interaction, a modified effective range expansion is possible where
\begin{multline}
\left[\frac{\Gamma(2l+2)}{2^l\Gamma(l+1)}\right]^2 [C_l(\eta_p)]^2 p^{2l+1}
(\cot\delta_l-i)=-\frac{1}{a_l}+\frac{1}{2} r_l p^2\\
+\frac{1}{4}s_l p^4+\cdots
-\frac{2k_C\, p^{2l}}{\Gamma(l+1)^2}
\frac{|\Gamma(l+1+i\eta_p)|^2}{|\Gamma(1+i\eta_p)|^2}H(\eta_p)\,,
\end{multline} and 
\begin{align}
C_l(z)&=\frac{2^l e^{-\pi z/2}|\Gamma(l+1+i z)|}
{\Gamma(2l+2)}\,,\nonumber\\
H(z)&=\psi(iz)+\frac{1}{2iz}-\ln(iz)\,,
\end{align}
with $\psi(z)$ the digamma function  and  $\kappa_C=\alpha_\text{EM}\mu$ the inverse Bohr radius. The $s$-wave phase shift for $l=0$ is then at the lowest order of approximation
\begin{align}\label{eq:ERE}
     C_0(\eta_p)^2 p(\cot\delta_0-i)&= -\frac{1}{a_0}-2\kappa_C H(\eta_p)\,,
\end{align}
 for small momentum $p$. The proton-proton scattering length $a_0=\SI{-7.81}{\femto\m}$~\cite{Piarulli:2015}. 

The phase shift in Eq.~(\ref{eq:ERE}) can be reproduced with a delta function potential $V_s(r)=c_0\delta(r)$ added to the long-range Coulomb $V_c(r)$. The numerical calculation of the phase shift, on a classical computer, can be done by discretizing the Schr{\"o}dinger equation on a spatial lattice with hard wall boundary condition in the relative separation of the two-particles such that $u(R_\text{wall})=0$ at some large radial distance $r=R_\text{wall}$ to be specified later in section~\ref{sec:results}. 

Near the origin $r=0$, the wave function $u(r)$ is regular. However, away from the origin, we have $u(r)/r\sim F_0(\eta_p, r p)+\tan\delta_0 G_0(\eta_p, r p)$ as a linear combination of the regular $F_0(\eta_p, r p)$ and irregular $G_0(\eta_p, r p)$ Coulomb wave functions, following standard derivations~\cite{Abramowitz,nist}. The hard wall at $r=R_\text{wall}$, then forces $u(R_\text{wall})=0$ 
which determines the phase shift as
\begin{align}\label{eq:phase}
    \delta_0(p) =\tan^{-1}\left[-\frac{F_0(\eta_p, R_\text{wall}\, p)}{G_0(\eta_p,R_\text{wall}\, p)}\right]\,,
\end{align}
where the momentum $p$ is obtained from the spectrum of the Hamiltonian on the lattice. The use of a hard outer wall to calculate phase shift for short-ranged interactions was introduced in Refs.~\cite{Carlson:1984,Borasoy:2007}.  
To reproduce the proton-proton phase shift described by the analytical result in Eq.~(\ref{eq:ERE}), we  tune the coupling $c_0$ which also regulates the divergence in the Coulomb potential $V_c(r)$ at the origin $r=0$. Such a calculation was carried out for a general 3-dimensional formulation in Refs.~\cite{RaviThesis, Rupak:2014xza} that we adapt to the 1-dimensional radial Schr{\"o}dinger equation here.  
In the 3-dimensional case, the spectrum of the Hamiltonian in Eq.~\ref{eq:3D_hamiltonian} without partial-wave expansion was regulated at the origin with  a $V(0)=c_0$ that was tuned to reproduce the $s$-wave phase shift~\cite{RaviThesis, Rupak:2014xza}.  A few results from the 3-dimensional classical lattice calculations at lattice spacing $b=\SI{1.97}{\femto\m}$ and lattice sites $L\gg R_\text{wall}=10$, 15, 22, 30,  40~\cite{Rupak:2014xza} are shown in Fig.~\ref{fig:DeltaIBMQ}.

The 1-dimensional calculation we perform here is similar to the 3-dimensional calculation discussed above. There are a couple of differences, however. A way to impose the boundary condition at the origin along with the short-ranged interaction $V_s(r)=c_0\delta(r)$ would be to put a wall at $r=0$, and a square-well potential over a small range to implement $c_0\delta(r)$. Numerically, we find regulating the  $V_s(r)=c_0\delta(r)$ and the repulsive Coulomb $V_c(r)=\alpha_{EM}/r$ at the origin works more accurately if we take  $V(0)=c_0$, $V(b)= [c_0+\alpha_{EM}/(2b)]/2$ and $V(r>b)=\alpha_\text{EM}/r$. 
This is the potential, with an offset $c_0$, for a charge uniformly distributed over a sphere of radius $2b$, and regulates the Coulomb divergence at $r=0$. The physics outside the radial distance $2b$ remains unchanged. 
The coupling $c_0$ that regulates the short-distance physics depends on the lattice spacing $b$ which is the ultraviolet lattice regulator but it is independent of the lattice size  $(L-1)b$ for $L\gg 1$. In Table~\ref{table:energyIBMQbrisbane} and Fig.~\ref{fig:DeltaIBMQ} we show several 1-dimensional calculations.  

\subsection{Relaxation method}
\label{subsec:relaxation}

The phase shift calculation from Eq.~(\ref{eq:phase}) needs the determination of the relative momentum $p$. In the classical lattice calculation~\cite{RaviThesis,Rupak:2014xza},  this was done from the lowest energy eigenvalue of the Hamiltonian. In quantum computation, the energy eigenvalues are determined from variational calculations starting from a trial wave function. In Ref.~\cite{Yusf:2024igb} we used the relaxation method~\cite{Schroeder:2017,Jackson:1998} to calculate the ground state energy of two particles in general coordinates interacting with a short-ranged force inside a harmonic trap. It is an iterative method leveraging quantum-classical calculations. Discretizing the Laplacian by nearest neighboring sites hopping $d^2 u(r)/dr^2=[u(r+b) +u(r+b)-2u(r)]/b^2$ suggests an iterative method where the updated value depends on the average of the nearest neighbors. This is a standard procedure for solving the Poisson's equation in electrostatic problems~\cite{Jackson:1998} since the solutions to the Laplacian are harmonic functions. We use 
\begin{align}\label{eq:relaxation}
   u^{(\text{new})}(r)= \frac{2 t\,\xbar{u}(r)}{2t+b^2[V(r)-E]}\, ,
\end{align}
iteratively where $E$ is the energy associated with the current wave function $u(r)$. $t=1/(2\mu)$ is the hopping parameter, and $\xbar{u}(r)$ is the nearest neighbor average. In the Jacobi method, the energy $E$ is calculated on a quantum computer and the wave function is updated 
$u(r)\rightarrow u^{(\text{new})}(r)$ iteratively on a classical computer until a desired rate of convergence is achieved. Alternatively, one can implement a Gauss-Siedel method~\cite{Schroeder:2017,Yusf:2024igb} where quantum computation is used to calculate the energy $E$ for the trial state. The subsequent updates are done classically.  We apply the latter method given the quantum computing resources that are available to us.

\section{2nd quantization and quantum computing}
\label{sec:qc}

The qubitization---formulation of the physical system on a quantum computer---is straightforward in the 2nd quantization language for the Hamiltonian in Eq.~(\ref{eq:radial_hamiltonian}). In 2nd quantization, space-time coordinates are treated as ordinary labels and instead fields defined at each space-time point are promoted to operators that satisfy anti-commutation relations to describe fermions.  
This naturally leads us to consider the problem in quantum field theory. We use an EFT where the short-ranged nuclear interaction is expanded systematically in the non-relativistic energy. 

The analytical calculation of the phase shift in EFT at leading order was detailed in Ref.~\cite{Kong-Randal:1999} with a follow-up calculation~\cite{Kong-Randal:1999B} that included the effective range $r_0$ contribution at next-to-leading order.  The  EFT Hamiltonian  for the system described by   
Eq.~(\ref{eq:radial_hamiltonian}) is 
\begin{align}
H=\int dr\, \Psi^\dagger(r)\left[-\frac{1}{2\mu}\frac{d^2}{dr^2}+V(r)\right] \Psi(r)\,,
\end{align}
which is written on a spatial lattice with $L$ sites as
\begin{multline}\label{eq:latticeH}
\hat{H}=-\hat{t}\sum_{l=0}^{L-2}\left[\hat{\Psi}_{l+1}^\dagger \hat{\Psi}_{l}
+\operatorname{h.c.}\right]+2\hat{t}\sum_{l=0}^{L-1}\hat{\Psi}_{l}^\dagger \hat{\Psi}_{l}\\ + \sum_{l=0}^{L-1}\hat{V}_l\hat{\Psi}_{l}^\dagger \hat{\Psi}_{l}\,,
\end{multline}
where dimensionful quantities are written in units of the lattice spacing $b$, and indicated with a hat. The second-derivative in the kinetic energy term has been approximated with nearest-neighbor finite differences. Open boundary conditions are used for the kinetic energy to impose the boundary conditions at $r=0,\, R_\text{wall}$.

The qubitization of the Hamiltonian in Eq.~(\ref{eq:latticeH}) for quantum computation requires identification of the second quantized fields ($\hat{\Psi}_{l}$, $\hat{\Psi}_{l}^\dagger$) with the qubits. We introduce a qubit $q_l$ for each lattice site $l$ that is used to indicate the presence $|q_l=1_l\rangle$ or absence $|q_l=0_l\rangle$ of a particle at that particular site, respectively. Fermionic degrees of freedom are mapped to qubits using the Jordan-Wigner transformation (JWT)~\cite{Jordan:1928}
\begin{align}\label{eq:JWT}
    \hat\psi_l\mapsto \frac{X_l+iY_l}{2}\prod_{k<l}Z_k\,,
\end{align}
though in this case the distinction between a boson and a fermion is not relevant since we are working in the relative coordinates of a two-particle non-relativistic system. The Pauli matrices $X$, $Y$, $Z$ at different sites, indicated by the subscript in Eq.~(\ref{eq:JWT}), commute. The Hilbert space of the system is represented by a tensor product of $L$ qubits forming a $2^L$ dimensional linear vector space. 
JWT then leads to the Hamiltonian
\begin{multline}\label{eq:hamQC}
\hat{H}_\text{QC}=2\hat{t} \mathbbm{1}-\hat{t}\sum_{l=0}^{L-2}\frac{X_{l+1}X_l+Y_{l+1}Y_l}{2}  \\
+\frac{1}{2}\sum_{l=0}^{L-1}\hat{V}_l\mathbbm{1}-\frac{1}{2}\sum_{l=0}^{L-1}\hat{V}_lZ_l\,,
\end{multline}
as a sum of tensor products of Pauli matrices. 

The calculation of the energy expectation value $E/b=\hat{E}=\langle\psi|\hat{H}_\text{QC}|\psi\rangle$ to be used in Eq.~(\ref{eq:relaxation}) is reduced to expectation value calculations of Pauli matrices where  the quantum state 
\begin{align}\label{eq:wavefunction}
|\psi\rangle=\sum_{l=0}^{L-1} \hat{u}_l |2^{l}\rangle\,,
\end{align}
corresponds to the single-particle wave function $u(r)$ at  $r=0,\dots, R_\text{wall}$ defined on the spatial lattice  as $\hat{u}_l$ for $l=0,\dots, L-1$. We recall that  $R_\text{wall}=(L-1)b$.  Arbitrary qubit states are usually labeled by the decimal representations $|k\rangle$  of the binary equivalent of zeros and ones indicating states of the qubits. Thus,   $|2^l\rangle$ represents the presence of a single particle at site $l$. The state 
$|\psi\rangle$ described above is a linear superposition of such states with probability amplitudes $ \hat{u}_l $ as appropriate. The Hilbert space of $L$ qubits is $2^{L}$ dimensional, out of which the non-relativistic wave function $|\psi\rangle$ contains only $L$ non-zero components. 

The expectation value $\langle\psi|\hat{H}_\text{QC}|\psi\rangle$ calculation in terms of the Pauli strings in Eq.~(\ref{eq:hamQC}) involves evaluations of $\langle \psi|Z_l|\psi\rangle$,  $\langle \psi|X_{l+1} X_l |\psi\rangle$ and  $\langle \psi|Y_{l+1} Y_l|\psi\rangle$. 
We calculate analytically the terms proportional to the identity matrix  as trivially $\langle\psi|\mathbbm{1}|\psi\rangle=1$ for normalized states $|\psi\rangle$. 

The expectation value $\langle\psi|\hat{O}|\psi\rangle$ of any operator $\hat{O}$ is just a sum of its eigenvalues $\lambda$ 
weighted by the corresponding probability $P(\lambda)$ for measurement. Thus,   $\langle \psi|Z_l|\psi\rangle= P(0_l)-P(1_l)$ where $P(q_l)$ is the probability to measure the $l$-th qubit with values $q_l=0,1$ since the qubits are quantized along the $\hat{z}$-direction. For measurements of the other two Pauli matrices, we transform to a basis where the Pauli matrices $X$ or $Y$ are diagonal as appropriate. Given that 
\begin{align}
X&= HZH^\dagger\,, & Y &=(S H^\dagger) Z (H S^\dagger)\,,\nonumber\\
H&=\frac{1}{\sqrt{2}}\begin{pmatrix} 1 &1\\ 1 &-1\end{pmatrix}\,, & S &= \begin{pmatrix} 1 &0\\ 0 &i\end{pmatrix}\,,
\end{align}
defining the qubit states $|\psi'\rangle= H_{l+1}H_l|\psi\rangle$ and $|\psi'\rangle= H_{l+1}S^\dagger_{l+1}H_l S^\dagger_l|\psi\rangle$, respectively, for measurements of $\langle \psi|X_{l+1} X_l |\psi\rangle$ and $\langle \psi|Y_{l+1} Y_l |\psi\rangle$ leads one to consider $\langle \psi'|Z_{l+1} Z_l |\psi'\rangle=P(0_{l+1}0_l)-P(0_{l+1}1_l)-P(1_{l+1}0_l)+P(1_{l+1}1_l)$. Thus, the energy expectation value calculation is reduced to single- and double-qubit measurements of $Z_l$ and $Z_{l+1} Z_l$, respectively, on the quantum computer. However, these measurements have to be done in the quantum superposition state of the complete wave function in Eq.~(\ref{eq:wavefunction}).

We make measurements with 10s of qubits. State $|\psi\rangle$ preparation on a quantum computer involves a large number of entangling two-qubit gates which leads to noisy measurements.  In Ref.~\cite{Yusf:2024igb}, with Rupak as a co-author, Schmidt decomposition was proposed for reducing state preparation to single- and double-qubit circuits.  We direct the reader there for details. The basic idea is that measurements of operators $Z_l$ and $Z_{l+1}Z_l$ involving single- and double-qubit measurements can be reduced to measurements in states with one and two qubits, respectively, by partial tracing over the subspace where no measurements are made. In particular, the Hilbert space is divided into subspace $A$ where the expectation value measurements of operator $\hat{O}_A$ are made, and subspace $B$ where no measurements take place. Then Schmidt decomposition gives~\cite{Yusf:2024igb}
\begin{align}
     \langle\psi| \hat{O}_A\otimes \mathbbm{1}_B |\psi\rangle&= \sum_{i=1}^{A}\lambda_i^2
     {}_A\langle i| \hat O_A|i\rangle_A\,,
\end{align}
 where the eigenvalues $\lambda_i^2$ and orthonormal basis states 
 $|i\rangle_A=\sum_a U_{ai}|a\rangle$ in subspace $A$ are obtained from the singular-value-decomposition (SVD)
 \begin{align}
 MM^\dagger = U \begin{pmatrix} \lambda_A^2 \\ &\lambda_{A-1}^2\\ & &\ddots \\ & & &\lambda_1^2\end{pmatrix} U^\dagger\,,
 \end{align}
 of the matrix $M$ constructed from the bipartite wave function
 \begin{align}
 |\psi\rangle\equiv \sum_{a,b} M_{ab}|a\rangle\otimes|b\rangle\,.
 \end{align}
 The probability amplitudes $M_{ab}$ are treated as matrix elements of $M$.  The matrix $M M^\dagger$ is of size $2\times2$ and $4\times4$ for single- and double-qubit measurements, respectively. 
 The trial state preparations in subspace $A$ after partial tracing requires a single two-qubit gate at most~\cite{Yusf:2024igb} for the two-qubit circuits, providing an enormous advantage in terms of error reduction.

\section{Results}
\label{sec:results}

\begin{table*}[htb]
\centering
\caption{IBMQ-Brisbane measurements were done with 5000 shots (quantum measurements)  with 
\texttt{optimization\_level=3}. The last five columns are energy measurements relative to the exact energy in the 4th column.  The ideal simulations were performed with 5000 shots as well, except for the simulations in column 8 that were done with 20000 shots, as indicated.}
\begin{ruledtabular}
\begin{tabular}{ccccccccc}
$L$ & $b$ (fm)  &$c_0$& Exact (MeV)& Full ideal &Full QPU
& SVD  ideal  & SVD  ideal (20000) &SVD QPU
\\ \hline \rule{0pt}{0.9\normalbaselineskip}
\begin{filecontents*}{IBMQ_Brisbane_table.csv}
L,a,czero,exact,fullideal,fullqpu,svdideal,svdidealB,svdqpu
15,2.0,-0.0449,5.809,9999,9999,1.090,1.038,1.323
20,2.0,-0.0449,5.809,9999,9999,1.122,1.022,1.464
25,2.0,-0.0449,5.809,9999,9999,1.120,1.032,1.407
7,1.0,-0.1112,22.13,0.9798,20.64,1.046,1.048,1.287
15,1.0,-0.1112,22.13,9999,9999,1.095,1.040,1.268
20,1.0,-0.1112,22.13,9999,9999,1.127,1.023,1.207
25,1.0,-0.1112,22.14,9999,9999,1.126,1.034,1.160
15,0.75,-0.1540,38.91,9999,9999,1.096,1.040,1.190
20,0.75,-0.1540,38.91,9999,9999,1.128,1.023,1.114
25,0.75,-0.1540,38.91,9999,9999,1.127,1.035,1.133
\end{filecontents*}
\csvreader[head to column names, late after line=\\]{IBMQ_Brisbane_table.csv}{}
{\   \L & \a &\czero
& \ifthenelse{\equal{\exact}{9999}}{---}{\num[parse-numbers=false]{\exact}}
& \ifthenelse{\equal{\fullideal}{9999}}{---}{\num[parse-numbers=false]{\fullideal}}
& \ifthenelse{\equal{\fullqpu}{9999}}{---}{\num[parse-numbers=false]{\fullqpu}}
& \ifthenelse{\equal{\svdideal}{9999}}{---}{\num[parse-numbers=false]{\svdideal}}
& \ifthenelse{\equal{\svdideal}{9999}}{---}{\num[parse-numbers=false]{\svdidealB}}
&\ifthenelse{\equal{\svdqpu}{9999}}{---}{\num[parse-numbers=false]{\svdqpu}}
}
\end{tabular}
\end{ruledtabular}
 \label{table:energyIBMQbrisbane}
\end{table*}  
The quantum computations in this work were performed on the QPU IBMQ-Brisbane (\verb|ibmq_brisbane|) available in the Open Plan. The calculations were done prior to July 2025 when the old cloud access was sunset. The results are shown in Table~\ref{table:energyIBMQbrisbane}. 

At a given lattice spacing $b$, we tune the coupling $c_0$ such that a numerical calculation of the phase shift using Eq.~(\ref{eq:phase}) from the ground state energy of the 1st quantized Hamiltonian in Eq.~(\ref{eq:radial_hamiltonian}) reproduces the analytical result in Eq.~(\ref{eq:ERE}). The same coupling is then used in formulating the 2nd quantized Hamiltonian in Eqs.~(\ref{eq:latticeH}) or (\ref{eq:hamQC}). We emphasize that it is not necessary to perform the calculations for tuning the couplings in the 1st quantized form. One can directly tune the couplings in the expectation value calculation from the measurements of the Pauli strings in Eq.~(\ref{eq:hamQC}) to obtain the correct phase shift. 

We perform quantum computations at $b= \SI{2}{\femto\m}$, $\SI{1}{\femto\m}$ and 
$\SI{0.75}{\femto\m}$ for various lattice sizes $L$, Table~\ref{table:energyIBMQbrisbane}. The energy expectations were calculated with a trial wave function of a particle in a box of size $5 b$, irrespective of the lattice size $L>5$. Through experimentation, we find that a more compact trial wave function yields less noisy results~\cite{Yusf:2024igb}. 

In Table~\ref{table:energyIBMQbrisbane}, the fourth column lists the exact numerical calculation of the trial state energy using matrix multiplications in the 1st quantized form. Results from both ideal simulations on Qiskit~\cite{qiskit2024} and QPU IBMQ-Brisbane are shown. The number of measurements for each Pauli string was 5000 except for the ideal simulation results in the 8th column that were obtained from 20,000 measurements as indicated. The error relative to the exact values (4th column) when comparing the ideal simulation results with 5000 (7th column) and 20,000 (8th column) measurements shows the expected factor of $\sqrt{5000/20000}=1/2$ improvement. 

Note that the energy expectations in Table~\ref{table:energyIBMQbrisbane} are nearly constant at different lattice sizes $L$ at a fixed lattice spacing $b$ since they use the same exact trial wave function with a support over 5 lattice sites.

The measurements in Table~\ref{table:energyIBMQbrisbane} were done in one- and two-qubit circuits using SVD as indicated. For $L=7$ and $b=\SI{1}{\femto\m}$, we also perform calculations with the full trial state $|\psi\rangle$ without SVD. This requires a 7 qubit state preparation that involves a large number of entangling gates and leads to substantial errors on the QPU, making it unsuitable for the relaxation method.

\begin{figure}[htb]
\begin{center}
\includegraphics[width=0.49\textwidth,clip=true]{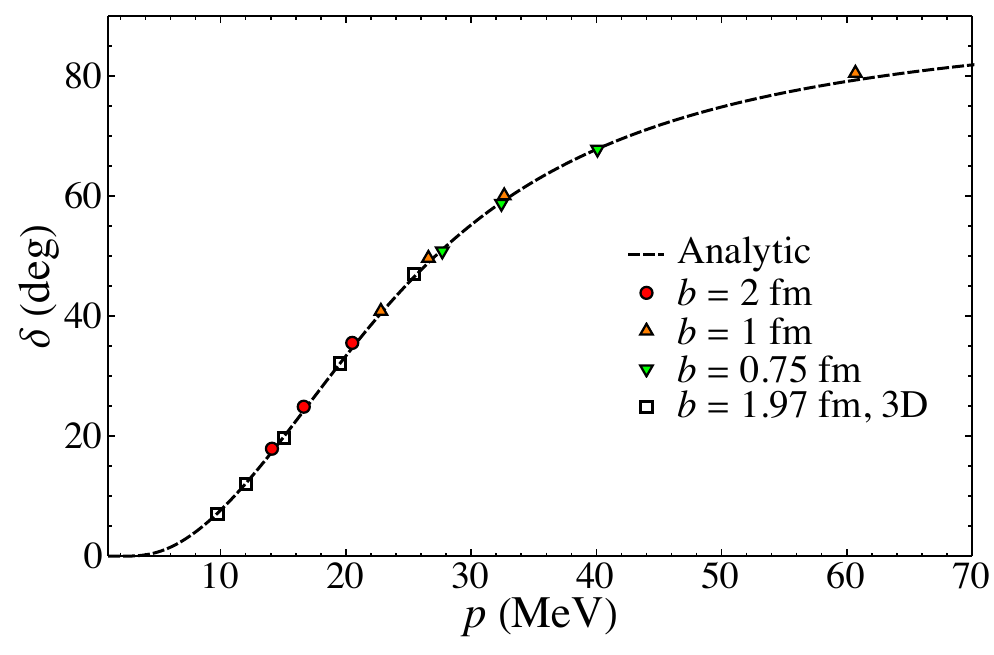}
\end{center}
\caption{\protect Proton-proton $s$-wave phase shift. The dashed (black) curve is the analytical result. The filled symbols are results from hybrid quantum-classical computation as described in the text. The open square symbols are results from a classical 3-dimensional  calculation~\cite{Rupak:2014xza}. Higher momentum points corresponds to smaller spherical wall size $R_\text{wall}$. }
\label{fig:DeltaIBMQ}
\end{figure}
In Fig.~\ref{fig:DeltaIBMQ}, the $s$-wave phase shift at several different lattice spacings and lattice sizes are compared with the analytical result from Eq.~(\ref{eq:ERE}). These phase shifts were obtained by using the trial energy expectation values from the SVD QPU measurements from Table~\ref{table:energyIBMQbrisbane} in the Gauss-Siedel relaxation method. After the initial QPU measurements, the rest of the relaxation method iterations are done classically. About 30 iterations provide sufficient accuracy, though we performed 200 iterations for these results.  Smaller physical box sizes $(L-1)b$ correspond to larger momentum. In the same figure, we show results from a 3-dimensional classical lattice calculation~\cite{Rupak:2014xza} for comparison. As discussed earlier, the QPU measurements were made with a particle in a box trial wave function. 
Other forms of trial wave functions in the variational calculations were tested in ideal and noisy Qiskit simulations,  and they gave very similar phase shifts.

We find accurate reproduction of the analytical phase shift from the quantum-classical calculation in Fig.~\ref{fig:DeltaIBMQ} over a range of lattice spacings $b$ and sizes $L$. The phase shift data near $p\sim \SI{60}{\mega\eV}$ for $b=\SI{1}{\femto\m}$ is slightly above the analytical result. This seems like a lattice discretization artifact. In ideal and noisy simulations at smaller lattice spacing $b=\SI{0.5}{\femto\m}$ but with the same physical volume $(L-1) b$, we find the accuracy improves.

\section{Conclusions}
\label{sec:conclusions}
We calculated the proton-proton elastic scattering $s$-wave phase shift at low momentum using a hybrid quantum-classical algorithm. The leading order  contribution is described by a short-ranged momentum independent two-particle interaction and the long-range Coulomb force. The system is formulated model-independently in the 2nd quantized form using EFT, which is used for the quantum computations.

The phase shift is calculated from the ground state energy of the system where a hard spherical wall is imposed in the relative separations of the protons. The boundary condition at the spherical wall for a given ground state energy directly leads to the phase shift from the relative contributions of the regular and irregular Coulomb wave functions.  

We use the relaxation method~\cite{Schroeder:2017,Jackson:1998} to obtain the ground state energy for the phase shift calculation.  This is an iterative variational calculation that provides an upper bound to the ground state energy. We implement the Gauss-Siedel formulation for the iterations where only an initial energy expectation value quantum computation is performed at the start of the process. The successive iterations do not require quantum computations~\cite{Schroeder:2017}, and can be performed classically.  The iterations for trial wave function updates can be performed using the Jacobi method as well, where each iterations would require a quantum computation of the energy expectation value. The IBM Open Plan we accessed did not have sufficient resources for the Jacobi method. Thus, we only implemented the Gauss-Siedel method on the QPU. 

 A straightforward implementation on the QPU requires trial state preparation involving 10s of qubits for our calculation. We use Schmidt decomposition to reduce all the calculations to at most two-qubit circuits~\cite{Yusf:2024igb}. This leads to a significant reduction in errors that varies between 10\% to 50\%. For this amount of error, the relaxation method is sufficiently robust, and converges to the exact ground state energy after about 30 iterations. State preparation of the full trial wave function without Schmidt decomposition leads to significant errors, and the relaxation method fails. 

The reduction of qubits for state preparation through Schmidt decomposition is fairly generic. It would impact other nuclear physics calculations where there is a hierarchy in the importance of the many-body forces, with the few-nucleon interactions being the most dominant. A recent work shows that at leading order chiral symmetry, the finite range of nuclear interactions, and nuclear saturation restrict interactions to at most four-nucleon forces in many-nucleon systems~\cite{Yang:2021vxa}.

\acknowledgments
The authors benefited from discussions with Samar Ghanam, Mark A. Novotny  and Muhammad Yusf.
This work was partially supported by U.S. DOE Grant No.  DE-SC0024286 and NSF Grant No. PHY-2209184.
We acknowledge the use of IBM Quantum services for this work. In this paper we used \verb|ibmq_brisbane| from the IBM Open Plan.

\bibliography{References.bib}
\end{document}